\documentclass[preprint]{aastex}

\newcommand{\degdot}{\mbox{$\hbox{$.\!\!^\circ$}$}}
\newcommand{\be}{\begin{equation}}
\newcommand{\ee}{\end{equation}}
\newcommand{\lab}[1]{\label{#1}}
\newcommand{\rr}[1]{~(\ref{#1})}

\begin{document}

\title{The Large-Scale Extinction Map of the Galactic Bulge from the MACHO
Project Photometry}

\author{Piotr Popowski} 
\affil{Max-Planck-Institut f\"{u}r Astrophysik,
Karl-Schwarzschild-Str. 1, Postfach 1317, 85741 Garching bei
M\"{u}nchen, Germany}
\affil{E-mail: popowski@mpa-garching.mpg.de}
\author{Kem H. Cook}
\affil{IGPP/LLNL, University of California, P.O.Box 808, Livermore, CA
94551, USA}
\affil{E-mail: kcook@igpp.ucllnl.org}
\author{Andrew C. Becker}
\affil{Bell Laboratories, Lucent Technologies, 600 Mountain Avenue,
Murray Hill, NJ 07974, USA}
\affil{E-mail: acbecker@physics.bell-labs.com}

\begin{abstract}
We present a $(V-R)$-based reddening map of about 43 square degrees of the 
Galactic bulge/bar. The map is constructed using template image
photometry from the MACHO microlensing survey, contains 9717
resolution elements, and is based
on $(V-R)$--color averages of the entire color-magnitude diagrams
(CMDs) in 4 by 4 arc-minute tiles. 
The conversion from the observed color to the reddening follows from
an assumption that CMDs of all bulge fields would look similar
in the absence of extinction. Consequently, the difference in observed
color between various fields originates from varying contribution
of the disk extinction summed along different lines of sight.
We check that our $(V-R)$ colors correlate very well with infrared
and optical reddening maps. We show that
a dusty disk obeying a ${\rm cosec} |b|$ extinction law,
$E(V-R) = 0.0274 \, {\rm cosec} |b|$, provides a good 
approximation to the extinction toward the MACHO bulge/bar fields.
The large-scale $(V-R)$-color and visual extinction map presented here
is publicly available in the electronic edition of the Journal and
on the World Wide Web. 

Subject Headings: dust, extinction --- Galaxy: center --- Hertzsprung-Russel diagram --- stars: statistics --- surveys
\end{abstract}

\section{Introduction}

There is a great interest in the study of stars in the central
Galactic region. Such investigations can answer questions about
the structure and extent of the Galactic bar, the metallicity gradient
in the inner Galaxy, the age of the dominant populations and their
formation history. These features can, in turn, provide crucial boundary 
conditions for cosmological simulations that attempt to
reproduce galaxies like the Milky Way at redshift $z=0$.

Most observations of the central Galaxy are still performed in the 
optical bands. 
Typical studies are carried out via two kinds
of surveys: (1) rather shallow large-scale studies aimed at detecting
spatial changes of stellar density, metallicity etc.; (2) deep
investigations of small fields aimed at detailed analysis of the local
stellar populations. Both treatments have to take into account
extinction. In the first case, the removal of reddening is essential for
reaching correct differential conclusions. In the second case, the fields
selected for observations are preferentially placed in low-extinction 
windows (e.g., Baade's Window), which allows detections of
intrinsically fainter
stars in observations with the same limiting magnitude. The selection
of new target fields, therefore, benefits from large-scale maps of 
extinction.

There are a few extinction maps of the central Galactic region.
The all-sky COBE/DIRBE map by Schlegel, Finkbeiner, \& Davis (1998)
covers this region, but there are suggestions that it overestimates 
the amount of dust in high-extinction areas (Stanek 1998; Arce \&
Goodman 1999; Dutra et al.\ 2003).
Schultheis et al.\ (1999) used DENIS infrared data to construct
a map of extinction in the Galactic plane with a latitude width
of about three degrees. 
The MOA Project prepared an extinction map of 18 square degrees
to determine the luminosity function for their analysis of
microlensing events (Sumi et al.\ 2003).
Dutra et al.\ (2003) investigated
extinction of the central $10^{\circ}$ of the Milky Way by analyzing 
the available 2MASS data. These maps covered many degrees of the sky.
Studies on smaller scales include Stanek's (1996) map in Baade's
Window and the Frogel, Tiede, \& Kuchinski (1999) map of several, 
small isolated
fields, located mostly at two perpendicular lines: one at the Galactic 
longitude $l = 0 \degdot 0$, and another at the Galactic 
latitude $b = -1 \degdot 3$.

Here, we use the optical $V$ and $R$ data (\S 2) from the MACHO experiment
(e.g., Alcock et al.\ 1995)
to construct a map of Galactic color with resolution of 4 arc-minutes.
The method and the resultant map are discussed in \S 3 and 4, respectively.
Our color map covers an area of about 43 square degrees and is based
on all of the 94 MACHO bulge fields ({\tt http://wwwmacho.mcmaster.ca/Systems/Coords/Fields.html}). 
This is a region that has not been fully covered by any other large-scale
extinction map, except for Schlegel et al.\ (1998), which may not
be reliable so close to the Galactic plane due to the presence
of contaminating sources. In \S 5 we argue that the geometry of the
dust distribution is well described by a very simple disk model.
In \S 6 we show that for the fields located at
the Galactic longitudes $l$ between 0 and 10 degrees and latitudes
$b$ between $-2$ and $-10$ degrees our color map may be converted
to an extinction map in a homogeneous fashion; three MACHO disk fields
at $(l,b) \approx (18^{\circ},-2^{\circ})$ require a different treatment.
This extinction map covers the Galactic latitude range that is
sparsely probed by other compilations. We summarize our results in \S 7.

\section{Data}

We use the photometric data from the MACHO microlensing experiment.
The MACHO experiment collected images of the Galactic bulge and
Magellanic Clouds from 1992 through 1999. All observations were taken with
the 1.3 m Great Melbourne Telescope with a dual-color wide-field camera.
The MACHO camera consisted of two sets of four 2k x 2k CCDs that
collected blue ($B_M$) and red ($R_M$) images simultaneously using
a dichroic beam splitter. A single observed field covered an area of
43' by 43'. Here we utilize a subset of the Galactic
bulge observations from 94 MACHO fields.

As we are not interested in light curves of the stars, but rather
in good representations of color-magnitude diagrams (CMD) of
different sky regions, we base our photometry on template images.
These are the images that were taken at the beginning of the MACHO
experiment during particularly clear, good seeing nights. Overall,
they provide the cleanest and most complete catalog of stars present
in the MACHO data.
Therefore, our CMDs are based on single-epoch photometry, but the
fraction of stars with highly variable color is negligible and will not
bias the mean colors we calculate. Each CMD is formed in a resolution
element referred to as a ``tile''. In MACHO terminology, tiles are
specific and unique 4 by 4 arc-minute areas on the sky.

The great majority of the MACHO bulge photometry has not been calibrated
and put on a uniform system. The only field that has a published 
calibration is MACHO field 119, which overlaps with Baade's Window.
To convert the MACHO instrumental magnitudes $B_M$ and $R_M$ to 
standard Johnson's $V$ and Kron-Cousins $R$, we use the following relations:
\be
V = 23.70 + 0.82 B_M + 0.18 R_M\label{v}
\ee
and
\be
R = 23.41 + 0.18 B_M + 0.82 R_M\label{r}.
\ee
Calibrations\rr{v} and\rr{r} were obtained from equations (1) and (2)
by Alcock et al.\ (1999), with $a_0$ and $b_0$ taken from their
section 5.1, $a_1$ and $b_1$ taken from their Table~2, average airmass
of about 1.04, and the bulge exposure time of 150s.
We apply these $V$ and $R$ calibrations derived
in field 119 to all our fields and do not make small-scale corrections
for the focal plane spatial irregularities (e.g., chunk corrections).
However, we apply global $(V-R)$ shifts to put all four
CCD chips (0 to 3) on the same system, with chip \# 1 serving as a
reference. These global shifts are given in Table~1.

\section{Our Method of Extinction Determination}

Interstellar extinction is typically determined by comparing the
measured color of an astronomical object (e.g., a star) with
the known (calibrated) intrinsic color.
The extinction can be estimated from two-band photometry using
a group of tracers rather than individual stars. This statistical
approach allows one to eliminate entirely erroneous measurements
that may result from the wrong classification or anomalous character
of an individual star. Stanek (1996) used the method of
Wo\'{z}niak \& Stanek (1996) utilizing red clump giants to construct
the map of extinction in Baade's Window centered at Galactic $(l,b) =
(1 \degdot 0,-3 \degdot 9)$. 
This map is based on the Optical
Gravitational Lensing Experiment (OGLE) observations in $V$ and $I$
standard filters and covers about half a square degree on the sky.
This map is intrinsically a map of relative extinction converting the
local surface density of stars to the expected amount of dust.
The currently accepted zero-point of extinction in this region
was estimated by Gould, Popowski, \& Terndrup (1998) and Alcock et
al.\ (1998). Kiraga, Paczy\'{n}ski, \& Stanek (1997) confirmed
the reddenings derived by Stanek (1996) from clump giants using not
only those stars but also two additional groups: subgiants and turnoff
stars. By comparing the results from all three groups of tracers,
Kiraga et al.\ (1997) showed that systematic errors of Stanek's (1996)
map are rather small.

There are two potential problems with Wo\'{z}niak \& Stanek's
(1996) approach.
First, for a map with high resolution (like 30'' by 30'' one by
Stanek 1996) there are generally few
stars in a single resolution element and the result is affected
by Poisson noise. Wo\'{z}niak and Stanek (1996) partially solved
this problem by ordering resolution elements according to the number
of stars with $V<20.0$ mag and then binning this series. This
procedure, however, introduces non-local correlations into a local 
estimate of extinction.
This is part of a more general second problem, namely the fact that
this procedure explicitly assumes that all the variations in stellar
surface density are due to extinction. This may be approximately true
in a small region like Baade's Window, but will not be correct if one
attempts to construct an extinction map of several square degrees.

Here, we propose a procedure that can circumvent the two above
problems.
First, it utilizes information stored in the whole set of observed
stars and not in just a limited group. Second, it makes no
assumptions about the underlying surface density of stars.
The method is based on the single assumption that the intrinsic
mean color of all observed stars (or their blends) in a given sky
region is always the same.
Of course, this assumption cannot be universally true. For example, we 
know that the CMD of the young disk population will be intrinsically noticeably
bluer than the CMD of an old bulge population. Consequently, we would
expect that the average of colors over all stars would be different.
We argue in \S 6.1 that despite the considerable extent of the MACHO
bulge region the first moment of the color distribution is dominated
by the bar stars and only marginally affected by the changing
contribution from the Galactic disk.
Nevertheless, some uncontrolled sensitivity to a particular population mix
in a given field remains the most serious systematic uncertainty
of the method.

For the MACHO experiment, the mean colors of CMDs
are very weakly dependent on the limiting magnitude
of the considered sample and, therefore, follow interstellar
extinction and not morphology changes.
This point is illustrated in Figure~1 which shows four representative
CMDs selected from almost ten thousand analyzed in this work. 
The CMDs were selected to differ in morphology; the differences arise
mostly from the various level of extinction.
The numbers in the right upper corners are the MACHO tile labels.
In each panel, the vertical solid line shows the average color of
the entire CMD.
We perform the following test. We divide each CMD into an upper and
lower part at two values of $V$ magnitude: 19.0, and 20.0.
Then we estimate the mean color of all four partial CMDs. The most
extreme values are plotted with a dotted lines in each of the
four considered CMDs. It is clear that a CMD color is very
insensitive to the part of the diagram that we select, and thus
the
method presented here is robust with respect to the level of extinction.
There are two reasons for this.
First, if the faint parts of a CMD are removed (which mimics
the effects of large extinction), the mean of color
stays the same.
Second, we find that the photometry in fields with large extinction
typically reaches to fainter apparent magnitudes (compare, for
example, T24653 and T48046). This is because
the number of recovered stars is to a large extent crowding and not 
magnitude-limited. As a result, CMD morphology in fields with different
extinction are more similar than one would naively expect.

The fact that
the CMD magnitude cut has little effect on the mean of the color
distribution is demonstrated in a statistical fashion in Figure~2,
which is made for all non-empty bulge tiles at our disposal.
The histograms presented in Figure~2 are centered close to 0.0 and
are relatively narrow, which shows that 
our CMD-based method is robust in the considered range of extinctions.

\section{The Map of Color}

To avoid random fluctuations of the mean CMD color due to 
small number statistics
and outliers, we eliminate all tiles with less than 1000 stars.
This leaves us with 9717 tiles with well-defined mean colors.
Most of the eliminated tiles are either entirely empty or only
partially filled with stars. This is a result of the original MACHO
procedure that divided the entire region in the sky into tiles, but
then placed field centers without any reference to tiles' positions.
As the scatter in $(V-R)$ for stars in a CMD is almost always below 0.3 mags
(and typically much smaller),
allowing only CMDs with at least 1000 stars will result in a Poisson
contribution of not more than 0.01 mag to the error in the average $(V-R)$.
Therefore, the total error in an average color, $<V-R>$, will be
dominated by the calibration errors.

The tiles that meet the above requirements are plotted in Figure~3. 
The color range is shown at the
top. The fields in the insert are MACHO disk fields that are spatially
separated from the others. We discuss them again in \S 6.
When these three fields at $(l,b) \approx (18^{\circ},-2)^{\circ}$ 
are removed the sample consists of 9422 tiles.

Two facts are immediately obvious from Figure~3:
\begin{enumerate}
\item the variation in observed color is so large that
differential extinction must be present since no population effects
can produce such large differences (see \S 6),
\item on large scales, color (extinction) is regularly stratified parallel
to the Galactic plane.
\end{enumerate}
We will discuss these features and their consequences in the next sections.

Figure~4 is a detail of Figure~3 that concentrates on the fields
closest to the Galactic center.  Contours of constant extinction from
Stanek (1996) are over-plotted on the Baade's Window region.  The
agreement is excellent.  We also mark several known and new
low-extinction windows.  These low-extinction areas are further
investigated in Figure~5 where we present $<V-R>$ versus Galactic
longitude $l$ in strips covering a range of $0 \degdot 4$ in Galactic
latitude $b$.  Low-extinction windows can be identified as troughs in
color that persist for at least two strips. A single possible
low-extinction field at $(l,b) = (3 \degdot 9,-3 \degdot 8)$ is seen
in only one $b$ strip, but the signal seems to be very strong.  Using
Figures~3, 4, and 5 we clearly identify all the currently known
low-extinction windows in the MACHO $(l,b)$ quadrant (Baade 1963;
Stanek 1998; Dutra, Santiago, \& Bica 2002). They are listed in
Table~2, where we also give the minimum $<V-R>$-color within
$0.25^{\circ}$ of the approximate field center. Note that Sgr~I window
(Baade 1963) is not very pronounced in our maps.  We also identify
three entirely new low-extinction windows at $(l,b)$ of $(2 \degdot
0,-3 \degdot 3)$, $(3 \degdot 2,-3 \degdot 4)$, and $(3 \degdot 9,-3
\degdot 8)$.  There is one possible low-extinction window somewhat
farther away from the Galactic center at $(3 \degdot 2,-6 \degdot
4)$. However, the detection of this window is based on color that is
uniformly bluer in one MACHO field. Thus it is not possible to
conclude whether the effect is due to lower extinction or a
particularly large calibration offset of this individual field.

\section{The Dusty Disk}

Observations of spiral galaxies (see NGC 891 for a beautiful example)
suggest that dust is typically concentrated in a thin disk located in
the plane of a galaxy. We will assume this geometry of the dust
distribution for the Milky Way, and see whether it leads to a
consistent picture. The simplest model of such a disk
would be one with a fixed vertical structure that is axisymmetric and
independent of the distance from the Galactic center. For a double
exponential disk such a model would have an infinite scale length.  If
we make an additional assumption that all the stars we observe are
behind the dust layer\footnote{As argued in \S 6.1, the assumption that
all the MACHO stars are behind the extinction layer is approximately
correct.} and that the Sun is located at the Galactic plane, then we
will get a familiar cosec law:
\be
E(V-R) = E_{\perp}(V-R) \, {\rm cosec} |b|,\lab{cosecb}
\ee
where $E_{\perp}(V-R)$ is an average color excess in the direction
perpendicular to the Galactic plane.  In this model, the observed mean
color is a sum of $E(V-R)$ and the intrinsic mean color of a CMD,
$<V-R>_0$,
\be
<V-R> = <V-R>_{0} + E_{\perp}(V-R) \, {\rm cosec} |b|. 
\lab{verysimple}
\ee

To avoid systematic effects associated with the analysis of different
populations, we exclude the MACHO fields that are entirely dominated
by the disk population: 301, 302, 303 at $(l,b) \approx 
(18^{\circ},-2^{\circ})$. However, we still retain
fields 304--311 at $(l,b) \approx (8^{\circ},-2^{\circ})$. Therefore, 
we perform our computation based on measurements for 9422 tiles.
The fit of equation\rr{verysimple} to the data is shown in the upper
panel of Figure~6. The lower panel shows the residuals of this fit.
The cone-like increase in the mean scatter with decreasing latitude
is not necessarily a problem as this type of behavior is expected
from more realistic disk models due to longitude dependence.
Ignoring the scatter issue, there is a systematic trend in residuals
suggesting that the model
can be improved. Unfortunately, most of the exponential models of the Galactic
disk share the problem apparent from Figure~6: it is very hard
to achieve a very fast change in color at low Galactic latitudes
and such a flat variation in the latitude range $(-6^{\circ},-10^{\circ})$.
Nevertheless, the model\rr{verysimple} is a good zeroth order approximation to
the data.
We consider two sources of error:
calibration errors in $<V-R>$ and Poisson errors in
$<V-R>$ originating from the number of stars in a given CMD.
We assume that the calibration error is identical
in all cases. Therefore, the total error, ${\sigma}_{\rm tot}$, is
given by

\be
{\sigma}^2_{\rm tot} = {\sigma}^2_{<V-R>,{\rm cal}} + {\sigma}^2_{<V-R>,P} \label{sigma2}
\ee
We obtain the calibration error in $<V-R>$ by fixing the chi square 
of the fit to equal the number of degrees of freedom, 
$\chi^2/d.o.f = 1$.
Using this procedure, we obtain $\sigma_{<V-R>,{\rm cal}} \approx
0.064$. This value is consistent
with the level of variation expected from the lack of a
field-by-field calibration, but most likely is only an upper limit
on the calibration error\footnote{Note that the uncertainty within
a single field is probably of order of 0.015 as can be judged from
the results in Baade's Window (field 119) described in \S 6.}.
The fact that model\rr{verysimple} is just a crude 
approximation to reality may be responsible for a large portion
of $\sigma_{<V-R>,{\rm cal}}$.
The best fit parameters for model\rr{verysimple} presented in
the form with uncorrelated errors are:
\be
<V-R> = (0.74156\pm 0.00066) + (0.02740 \pm 0.00012)\, ({\rm cosec}
|b|-{\rm cosec} |-4 \degdot 6513|).
\lab{dustydiskfit}
\ee
The very small errors in equation\rr{dustydiskfit} are internal
errors. The systematic errors 
coming from uncertainty in the global zero points of photometry are much
larger.

To obtain the results reported above we excluded three disk fields
at $(l,b) \approx (18^{\circ},-2^{\circ})$. 
However, fit\rr{verysimple} may be extended to include two intrinsic
colors, one for the bulk of the fields and the additional one just for fields
301, 302, and 303. The resultant $<V-R>_0$ for the high-$l$ fields
is $0.15$ magnitudes bluer, which may be explained by our theoretical arguments
from \S 6.1.

We would like to warn the reader against extensive use of 
eq.\rr{dustydiskfit}, since it provides a good representation only in
the global sense. On the other hand, all the local features of
reddening must be described by the detailed maps, like the one
presented here.

\section{Toward the Map of Extinction}

\subsection{General considerations}
If we accept the conjecture that the color excess is associated with
extinction, it is clear that we may bias the extinction measurement
by an incorrect assumption about the intrinsic color.
The fields that are dominated by the disk population should have
bluer intrinsic color, and as a result higher extinction than the
bulge-dominated fields with the same observed $<V-R>$.
In addition, some foreground disk stars will not be affected by
extinction, which would bias the observed mean color bluer and, as a
result, artificially lower the inferred extinction.
Here we estimate the biases caused by these effects.

Alcock et al.\ (2000) argue that in the MACHO bulge fields,
the disk contribution to the number
of stars down to $V = 23$ is around 10\%. Let us assume the same
level of disk contribution for the MACHO stars considered here which
were recovered with standard,
point spread function photometry (with completeness dropping steeply
at V = 21.5).
Let us consider two types of CMDs: one entirely dominated by bulge
stars with the intrinsic color $<V-R>_{0,B}$ and another one with
$\xi_d = 0.1$ disk contribution. If the intrinsic color of the pure disk population
is  $<V-R>_{0,D} \equiv <V-R>_{0,B} + \Delta_{DB}$ then the intrinsic
color of the second type of CMD will be 
$<V-R>_{0,B} + \xi_d\, \Delta_{DB}$. We will take $\Delta_{DB} = -0.3$
indicating that the disk population is on average 0.3 mags bluer
in $<V-R>$ than the bulge population. This is likely an upper limit
as can be judged based on integrated colors of populations with
different ages and metallicities (Girardi et al.\ 2002:
the tables provided on their web page: {\tt http://pleiadi.pd.astro.it}).
This would suggest that a CMD entirely dominated by bulge
stars has an intrinsic color that is only 0.03 mags redder than a
CMD with a 10\%  disk contribution.

Now, let us estimate how the presence of foreground disk stars
influences the inferred mean color and so biases the predicted
extinction low. 
We want to estimate what fraction of the disk stars are affected
by the total extinction and what fraction are less reddened.
To obtain qualitative results we will assume that the scale height
of the dusty disk is about 100 pc. In addition, we will make a
simplifying assumption that all stars that are more than 100 pc from
the plane experience full extinction and the ones that are less than
100 pc from the plane are foreground stars and experience no
extinction. (These are very
conservative assumptions with respect to estimating the effect of
foreground stars on the CMD color since there are no stars that can avoid
extinction entirely.)
The fields closest to the Galactic plane (for the MACHO coverage,
these are at $b \approx -2^{\circ}$) will be affected to the
highest degree.
For $b=-2^{\circ}$, the line of sight will be more than 100 pc from
the plane in somewhat less than about 3 kpc, which is about
a third of the distance toward the Galactic center.
For a typical MACHO field location and a typical stellar disk
model of the Milky Way, the number of disk stars observed in a given field
is a weak function of the distance along the line of sight 
(e.g., Kiraga \& Paczy\'{n}ski 1994).
Therefore, in our simple model $f = 1/3$ of the disk stars are
in front and $(1-f) = 2/3$ behind the extinction layer.
Consequently, 1/3 of disk stars or 3\% of all stars in such a
CMD are extinction-free and the rest are reddened by the same amount
independent of whether they belong to the disk or to the bar.
Given $E_{\rm true}(V-R)$, we estimate the observed color excess 
of the mixed population to be equal to
$E_{\rm obs}(V-R) = f \cdot \xi_d \cdot 0.0 + ((1-\xi_d) +
(1-f) \cdot \xi_d) \cdot E_{\rm true}(V-R) = 
(1-f \cdot \xi_d) \cdot E_{\rm true}(V-R)$.
For a representative $E_{\rm true}(V-R) = 0.5$, $E_{\rm obs}(V-R) =
0.483$, so that the bias amounts to only $0.483- 0.500 = -0.017$ mags.

For the variation of the disk contribution expected in the MACHO bulge
fields, both biases are negligible compared to the color excess itself.
However, it can be seen from the above arguments that if the disk
contribution is substantial, the difference in $(V-R)$-color excess
between a given CMD and a pure bulge CMD
can exceed 0.1 mags.
We believe that this is exactly the reason why the disk fields
presented in the insert in Figure~3 have systematically
bluer color than their low-$l$ counterparts at similar $b$ values. 
Therefore, the fields dominated by the disk
stars cannot be treated with the same extinction formulae as the bulk
of the MACHO fields. That was also the motivation to exclude some of high
longitude fields in the derivation of the dusty disk model in \S 5.

The MACHO bulge fields are situated in front of the tidal debris from the 
Sgr dwarf galaxy. The varying contribution of the Sgr Dsph at different
locations could bias our results as well. This is not a serious
concern at low Galactic latitudes as Sgr
contribution to the number of stars, as traced by the RR Lyrae population,
is on average less than 3\% in these fields (Alcock et al.\ 1997). 
Therefore, the bias caused by variation
in the number of Sgr stars should be substantially smaller
than the contribution from the Galactic disk and so completely
negligible.
The situation is much harder to quantify in the MACHO fields that are
farther away from the plane. Our investigation of RR Lyrae location
and extinction (Kunder, Popowski, \& Cook 2003, in preparation) will shed
more light on this problem.

\subsection{Extinction calibration through comparison with other
reddening maps}

Figure~7 confirms that our interpretation of variation in mean color
as due to differential extinction is correct. The left panel displays
the relation between the mean color determined in this paper
and extinctions obtained by Stanek (1996) for Baade's Window.
As the resolution of our map is lower than that of the OGLE-based
extinction map in
Baade's Window, each $A_V$ value for comparison was obtained by 
averaging between 64 and 81 of Stanek's (1996) 30'' by 30'' resolution 
elements.
The 82 points in Figure~6 show a clear correlation between the visual
extinction, $A_V$, and the mean stellar color, $<V-R>$.
The middle panel presents the comparison with Dutra et al.\ (2003)
values, that were based on the analysis of giant branches from 2MASS
data. Here we plot their $A_K$ extinctions versus our $<V-R>$.
We found the corresponding Dutra et al.\ (2003) values for about 1/9
of our resolution elements by selecting matched pairs on a one-to-one
correspondence basis. This procedure resulted in 1069 matches, which
span a much larger range in $<V-R>$ than the ones from
Stanek (1996).
Finally, in the right panel we plot Schlegel et al.\ (1998) reddenings
obtained by feeding their software with the positions of our
resolution elements. Since the Schlegel et al.\ (1998) reddening map
covers the entire sky, we matched all 9422 tiles.
The correlation between our color map and all the overlapping
extinction maps is very strong. Therefore, our $(V-R)$-color map
is an excellent map of {\em relative} reddening. For many
practical applications one is interested in going one step further
and inferring the value
of extinction based on the observed color. 
We use Stanek (1996) and Dutra et al.\ (2003) data to derive
the relations between the visual extinction $A_V$ and $<V-R>$.
We do not make fits using Schlegel et al.\ (1998) map because
$E(B-V)_{\rm SFR}$ versus $<V-R>$ relation cannot be universally
fitted with a straight line.
We discuss the calibrations based on maps by Stanek (1996) and Dutra et
al.\ (2003) separately.

\subsubsection{Extinction calibration based on Stanek's (1996) map}

We treat $<V-R>$ as an independent variable and fit a straight line 
of the following form:
\be
A_V = a + b <V-R>. \lab{teofit}
\ee
The most straight forward interpretation of the fit parameters would
imply that $b$ coincides with the coefficient of selective extinction
and $a$ is a product of this coefficient and an intrinsic
CMD color taken with a negative sign ($b \equiv R_{V,VR}$, 
$a \equiv -R_{V,VR}\, <V-R>_0$).
We consider three
sources of errors: calibration errors in $<V-R>$, Poisson errors in
$<V-R>$ originating from the number of stars in a given CMD, and
Poisson errors in $A_V$ coming from the scatter in Stanek's (1996)
extinctions.
We assume that our calibration errors are identical
in all cases. 
We assign all errors to the dependent variable.

Therefore, the total error, ${\sigma}_{\rm totBW}$, is
given by
\be
{\sigma}^2_{\rm totBW} = {\sigma}^2_{A_V} + b^2 \left(
{\sigma}^2_{<V-R>,P} + {\sigma}^2_{<V-R>,{\rm calBW}} \right) \label{sigma1}
\ee
We obtain the calibration error of $<V-R>$ by fixing the chi square 
of the fit to equal the number of degrees of freedom, 
$\chi^2/d.o.f = 1$. 
Application of this procedure to the data from Figure~6 results 
in ${\sigma}_{<V-R>,{\rm calBW}} = 0.0157$, which dominates over the
other sources of noise.
If the calibration noise is Gaussian, then the expected number
of points deviating from the fit by more than 2.75 sigma is $\approx
0.5$. Therefore, we will treat all points that deviate by more than
2.75 sigma as outliers, remove them, and repeat the fit.
This procedure should make our results less
prone to systematics.
The one outlier is found at $2.98\sigma$. When removed,
the $\chi^2$-normalized scatter drops to $\sigma_{<V-R>,{\rm calBW}} =
0.0146$. 
We obtain the following fit presented in
the form with uncorrelated errors :
\be
A_V = (1.595 \pm 0.006) + (3.52 \pm 0.12)\left[ <V-R> - 0.7861 \right]
\lab{bestfitav}
\ee
Relation\rr{bestfitav} is shown in Figure~7 as a dashed line. The
circled point marks the excluded outlier.

There are three potential problems with solution\rr{bestfitav}:
\begin{enumerate}
\item all points with $<V-R>$ larger than 0.86 are above the fitted
curve, which suggests some systematic trend;
\item the slope of $3.52 \pm 0.12$ is much lower than the standard 
coefficient of selective extinction $R_{V,VR} \sim 5$ [The extinction 
toward the Galactic bulge may be anomalous as discussed by Popowski
(2000) and Udalski (2003), but the value obtained here is rather
extreme.];
\item zero extinction is reached only for $<V-R>_0 = 0.33 \pm 0.02$,
which is probably bluer than the colors of the bulge turn-off stars that
determine the mean CMD color.
\end{enumerate}

Note that equation\rr{dustydiskfit} from \S 5
implies $<V-R>_{0} = 0.4037 \pm 0.0016$, which is $0.07$ magnitudes
redder than $<V-R>_{0} = 0.33 \pm 0.02$ suggested by fit\rr{bestfitav}.
This mismatch can originate from several sources.
First of all, the zero point of Stanek's (1996) extinction map
can be overestimated by about $0.25$ mags. This is not very likely
as this zero point has been accurately measured with two independent types
of stars (Alcock et al.\ 1998; Gould et al.\ 1998).

Second, relation\rr{verysimple} is from the operational point of view
indistinguishable from
\be
<V-R> = <V-R>_{\rm const} + E_{\perp}(V-R) \, {\rm cosec} |b|, 
\lab{verysimpleconst}
\ee
where $<V-R>_{\rm const}$ can have two non-zero components:
the intrinsic mean color of the CMD, $<V-R>_0$, and the uniform
over-the-region component of extinction. In general, external information 
about the uniform extinction or intrinsic $<V-R>_0$ color of the
bulge population would be needed to make a unique decomposition of
$<V-R>_{\rm const}$.
The color mismatch would be explained if a uniform extinction screen
covered most of the MACHO Galactic fields with exception of Baade's
Window (and maybe few others).
Two sources of possible uniform extinction: zodiacal dust and the Local
Bubble are ineffective in producing significant extinction sheets.
Schlegel et al.\ (1998) argue that the light
to dust ratio for the material contributing to the zodiacal light is orders
of magnitude larger than for ordinary interstellar matter. They estimate
that zodiacal extinction is at the level of $10^{-6}$ mags in optical
filters.
The possible maximum contribution of the idealized bubble and the real 
geometry of the 
Local Bubble argue against a uniform layer of extinction of the
required magnitude.
Therefore, it is justified to assume that the constant term in 
equation\rr{verysimpleconst} coincides with intrinsic color, that is
$<V-R>_{\rm const} \approx <V-R>_0$.

This leaves us with the third option, namely, that some systematics in
Baade's Window chooses a $\chi^2$ minimum that is not a global
minimum for all MACHO fields. This conclusion is supported by the very
blue $<V-R>_0$ and anomalously small $R_{V,VR}$ reported in 
equation\rr{bestfitav}.
When we fix $<V-R>_0 = 0.4037$, fit\rr{teofit} to the Baade's Window
data from Figure~7 results in $R_{V,VR} = 4.17 \pm 0.01$ (solid line
in the left panel).
We believe that the above values are preferred over
the ones returned by unconstrained fit\rr{bestfitav}, but the final
resolution of the mismatch in intrinsic color will be addressed
with a few thousand RR Lyrae stars (Kunder et al.\ 2003, in preparation)
selected from the MACHO database.

\subsubsection{Extinction calibration based on Dutra et al.\ (2003) map}

We follow the route very similar to the one outlined in \S 6.2.1.
To avoid any assumptions about the selective extinction coefficient,
we try to recover the raw results from Dutra et al.\ (2003).
Therefore, we convert their $A_K$ extinctions to
$E(J-K_s)_{\rm 2MASS}$ dividing all $A_K$ values by a factor of 0.670,
which they used.
For convenience, we then switch to $E(J-K)_{UKIRT}$ using color
transformations given in
{\tt http://www.ipac.caltech.edu/2mass/releases/allsky/doc/sec6\_4b.html}.
We scale the errors in the same fashion.
We derive the following fit presented in the form with uncorrelated
errors:
\be
E(J-K)_{\rm UKIRT} = (0.2430 \pm 0.0024) + (0.657 \pm 0.016)
(<V-R>-0.839). \lab{jmkfit}
\ee
We detect no outliers.
Appropriately scaled line\rr{jmkfit} is shown in the middle panel of
Figure~7.
Result\rr{jmkfit} is very insensitive to the level of the assumed MACHO
field-to-field calibration errors, ${\sigma}_{<V-R>,{\rm cal}}$.
The fit parameters change by less than 0.1\% for ${\sigma}_{<V-R>,{\rm
cal}}$ values between 0.02 and 0.064 mags, where the upper limit comes
from the scatter derived during the analysis of the disk model in \S
5. To be specific, equation\rr{jmkfit} was derived using 
${\sigma}_{<V-R>,{\rm cal}} =0.02$ mag.
We believe that the errors in the field-to-field calibration cannot
be smaller than 0.02 mags. However even for this very low value, our fit
results in $\chi^2/d.o.f. = 0.167$. We conclude that Dutra et al.\ (2003)
errors are too conservative. They are likely to be overestimated
on average by at least a factor of 2.8.

For the one-parameter extinction curves from Cardelli, Clayton, \& Mathis
(1989) and O'Donnell (1994), the slope from
equation\rr{jmkfit} implies that $R_{V,VR} = 4.56 \pm 0.05$.
Therefore, the visual extinction $A_V$ is given by:
\be
A_V = -2.14  + 4.56 \, <V-R>. \lab{avfrom2mass}
\ee
Equation\rr{avfrom2mass} yields the intrinsic CMD color
of $<V-R>_0 = 0.47$,
substantially redder than what we derived from the Stanek (1996) map.
Here, we do not attempt to fix the zero point to the value obtained
from the dusty disk model, because our tiles matched to Dutra et al.\
(2003) resolution elements span wide enough range in $<V-R>$ to
produce robust results.

Finally, we note that the total visual extinction out of the plane
can be expressed as $A_{V,\perp} = R_{V,VR} \, E_{\perp}(V-R)$. Let us 
assume that $R_{V,VR} = 4.4$, a representative value based on 
calibrations to Stanek (1996) and Dutra et al.\ (2003) extinctions.
For $E_{\perp}(V-R) = 0.0274$ taken from eq.\rr{dustydiskfit}, 
out-of-the-plane extinction  $A_{V,\perp} = 0.12$.
It is interesting that this value obtained from observations close
to the Galactic plane is consistent with results from extinction studies
investigating the surroundings of the North Galactic Pole (e.g., Knude
1996).

\section{Summary}

We have constructed a $(V-R)$-color map of about 43 square degrees 
of the bulge/bar central region. This map has a resolution of 4' by 4' and
is based on photometry from the MACHO microlensing survey.
Each resolution element is assigned a color that is an average color
of all stars detected in a tile from the MACHO template image. We selected
only tiles with at least 1000 stars, and a typical tile contains
a few thousand stars. 
We argued that average colors are insensitive to the details of the
CMD morphology, which makes them a robust tool in the considered
extinction range.
We presented evidence that even a very simplistic cosec-law model
of the Galactic dusty disk can qualitatively explain the observed
distribution of mean color. However, we encourage the reader to use
extinction maps whenever available, because the smooth disk model
does not provide a good local approximation.
We think that the geometry of the extinction layer revealed by the
${\rm cosec} \, |b|$ fit in Section 5 and observations of external 
galaxies argue strongly against the large amount of dust in the Galactic bulge
itself.  Even if an internal bulge extinction were present, the method
implemented here  would still work properly, because it simply
measures the extinction toward the dominant population represented in
a given CMD. However, the interpretation of the extinction as merely a
mean value would become more important.  At the moment, we interpret
our reddenings as universally applicable to most stars observed toward
the bulge lines of sight (except for a small number of foreground disk
stars). Substantial internal reddening in the bulge would cause
closer {\em bulge} stars to have overestimated reddenings and far ones to
have underestimated reddenings. There would be no easy way to avoid
this additional source of scatter.

Comparison with extinction maps by Stanek (1996), Schlegel et al.\
(1998), and Dutra et al.\ (2003)
showed that the mean CMD colors correlate strongly with extinctions.
Therefore, our color map can serve as an extinction map.
Discussions in \S 5 and \S 6 suggest that there exists some
uncertainty in fit parameters needed to convert CMD colors to
visual extinctions. 
This question is being 
currently addressed through comparison of this color map with color
excesses of RR Lyrae stars (Kunder et al.\ 2003, in preparation)
selected from the MACHO database.
For the time being, we give two possible prescription for
obtaining $A_V$ from $<V-R>$:
\be
A_V = -1.68 + 4.17\, <V-R>. \lab{extconv1}
\ee
based on Stanek's (1996) map and:
\be
A_V = -2.14 + 4.56\, <V-R>. \lab{extconv2}
\ee
based on the extinction determination by Dutra et al.\ (2003).
The extinctions from the second calibration [eq.\rr{extconv2}]
are about 0.2 mags lower for our weakly reddened fields
with $<V-R> \sim 0.6$. The agreement is much better for
$<V-R> \in (1.0,1.4)$. 
A complete extinction table will be published in the electronic
version of the Journal.
Table 3 provided here illustrates its format and content.
The five columns list the positions
in Galactic $(l,b)$ coordinates, $<V-R>$ - color, and two values
of extinction calibrated according to equations\rr{extconv1} 
and\rr{extconv2}. The uncertainty in extinction 
comes from two main sources: a mismatch
between calibrations\rr{extconv1} and\rr{extconv2} discussed above
and field-to-field calibration errors in $<V-R>$.
When propagated to $A_V$, the field-to-field calibration errors are 
likely not smaller
than $4.17 \times 0.02$ and $4.56 \times 0.02$ mags for the values
derived from equations\rr{extconv1} and\rr{extconv2}, respectively.
We suggest using $\sigma_{A_V} = 0.1$ mag as a lower limit
in both cases\footnote{The uncertainty associated with a mismatch between 
calibrations\rr{extconv1} and\rr{extconv2} must be dealt with
separately and its importance depends on the $<V-R>$ range.}.
At the end of the electronic Table 3, we also include extinction
estimates for the three disk fields at 
$(l,b) \approx (18^{\circ},-2^{\circ})$. We account for population
effects in those fields by substituting $<V-R>$ with $(<V-R> + 0.15)$ 
in equations\rr{extconv1} and\rr{extconv2}.
The color and extinction maps presented here are also available from 
{\tt http://www.mpa-garching.mpg.de/~popowski/bulgeColorMap.dat}.

\acknowledgments

Kem Cook thanks Max-Planck-Institut f\"{u}r Astrophysik, where
part of this work was completed, for its hospitality.
This work was performed under the auspices of the US Department
of Energy, National Nuclear Security Administration, by the University
of California, Lawrence Livermore National Laboratory, under
contract W-7405-ENG-48.

\clearpage

\begin{figure}[htb]
\includegraphics[width=16cm]{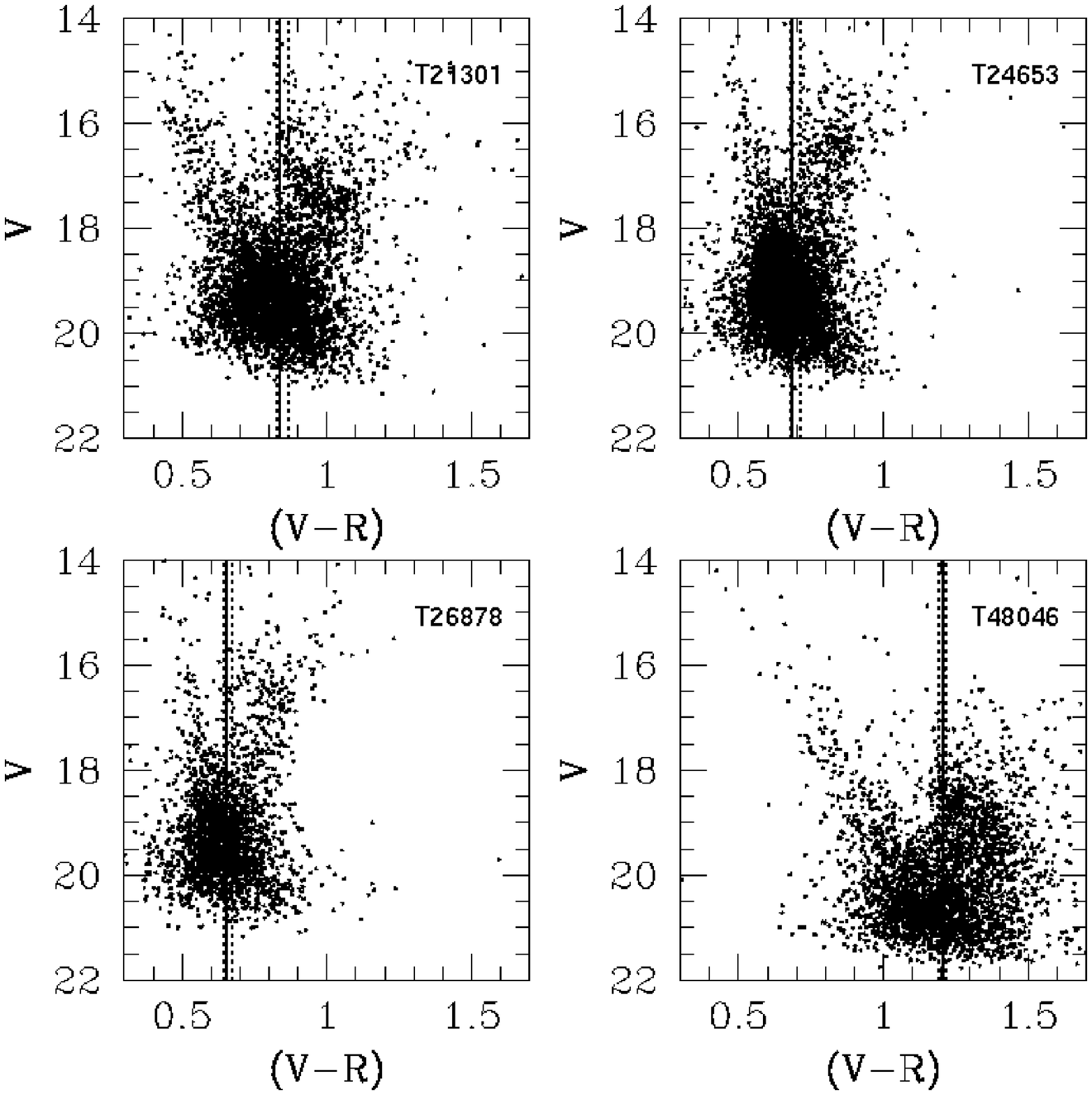}
\caption{Four representative CMDs from the MACHO database.
Tile names are given in the upper right corners. The solid line marks
the average color. The dotted lines indicate the most extreme average
colors obtained from analyzing partial CMDs containing stars either
brighter of fainter than $V=19.0, 20.0$. It is evident that the mean
color is a very stable quantity with respect to simulated extinction bias.
\label{figure1}}
\end{figure}

\begin{figure}[htb]
\includegraphics[width=16cm]{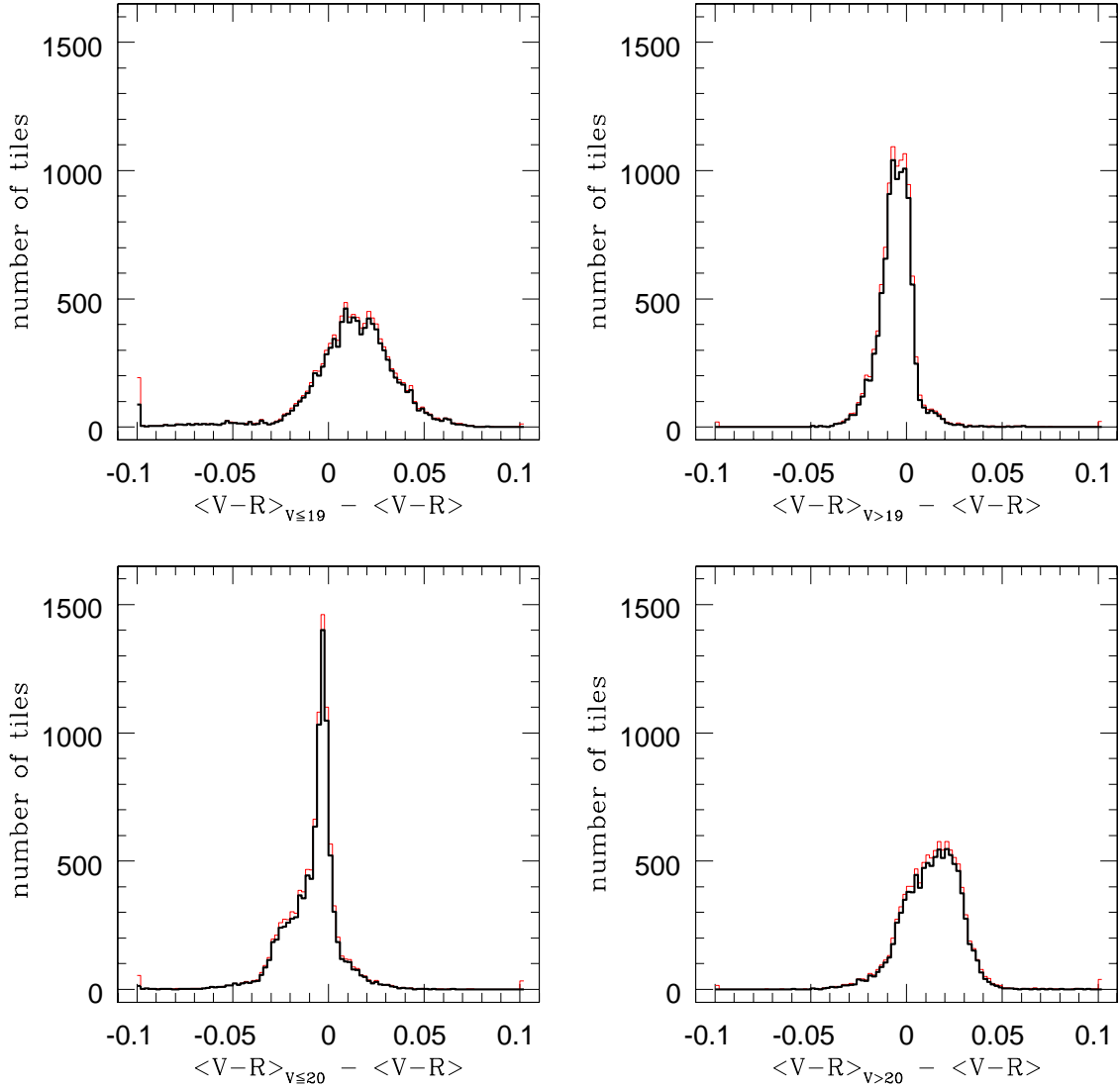}
\caption{Histograms of the difference between average colors as
computed from the entire MACHO CMD for a tile and the CMD cut at a certain $V$
magnitude, i.e., partial CMDs containing stars either brighter 
or fainter than $V = 19.0, 20.0$. In all cases, the difference is
small and its distribution
very narrow. The thin (red) line represents all 10466 non-empty tiles and
the thick (black) line correspond to the tiles with more than 1000
stars.
\label{figure2}}
\end{figure}

\begin{figure}[htb]
\includegraphics[width=16cm]{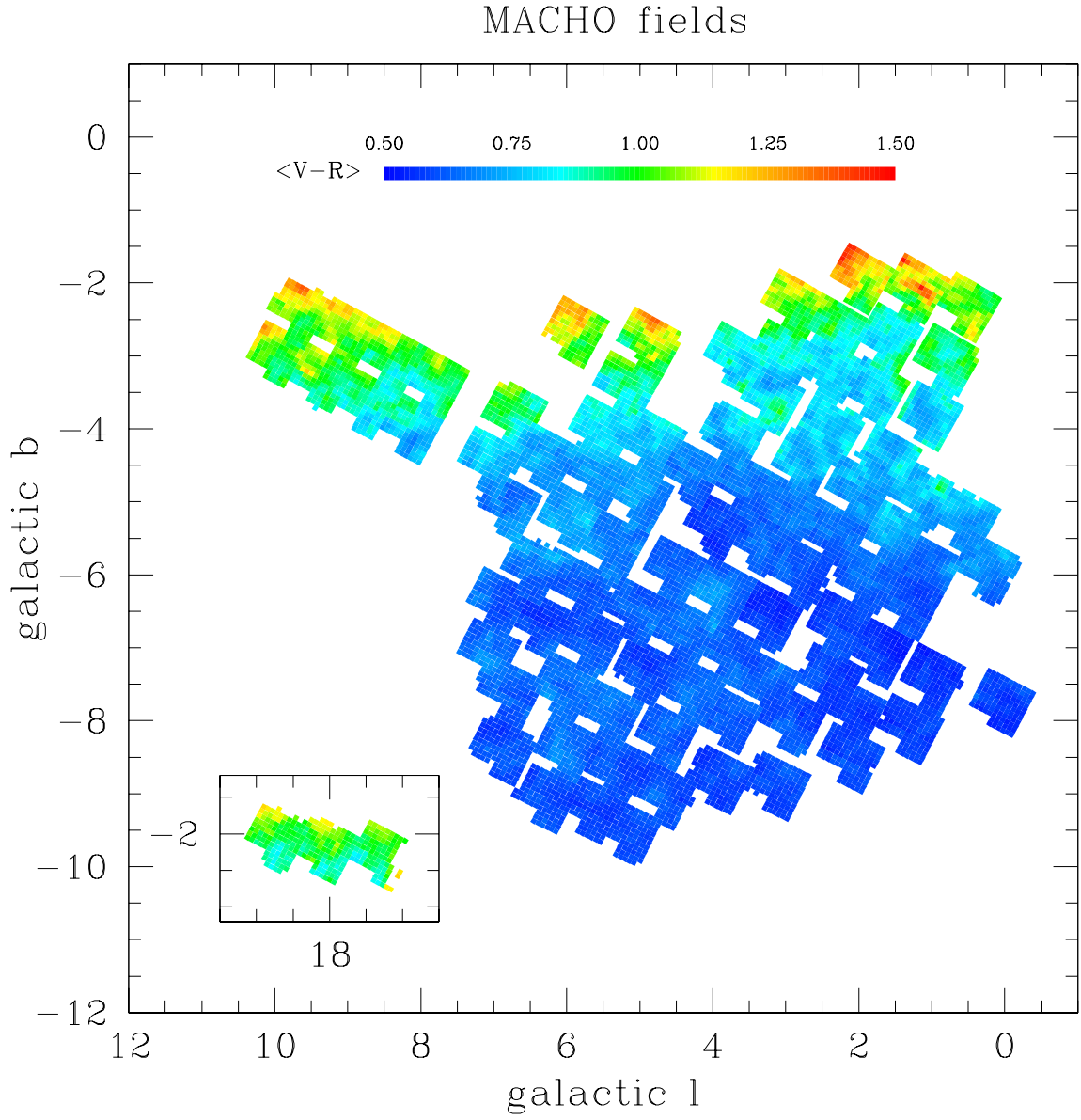}
\caption{Color map of the central Galactic region. The scale is given
at the top. Insert presents three disk fields separated from the others by
$10^{\circ}$. The entire map is composed of 9717 resolution elements,
4' by 4' each.
\label{figure3}}
\end{figure}

\begin{figure}[htb]
\includegraphics[width=16cm]{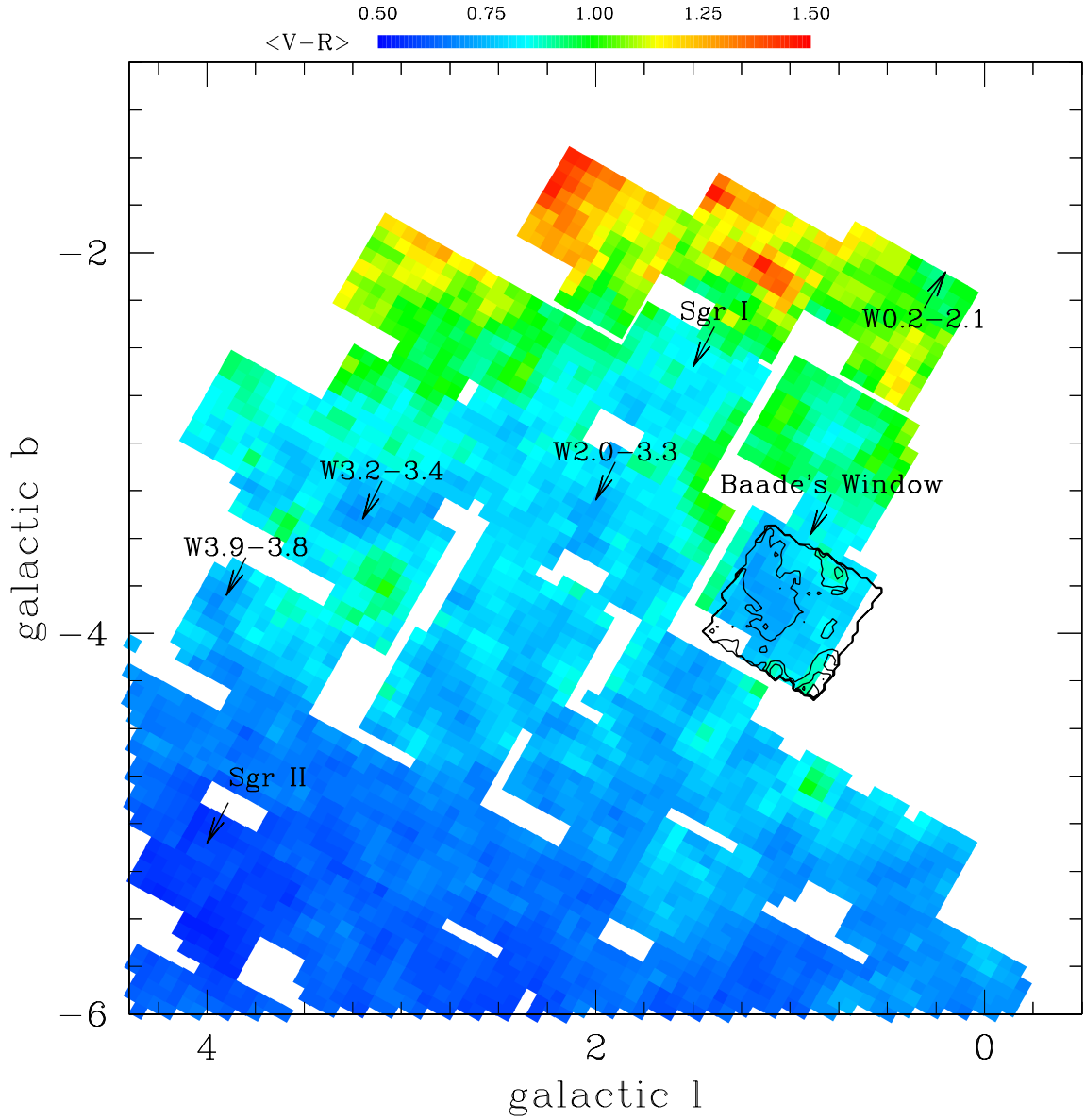}
\caption{Color map showing details of the central Galactic
region from Figure~3. The scale is given at the top.
Stanek's (1996) extinction contours are over-plotted on our Baade's
Window map. Additional low-extinction windows are marked with black
arrows.
\label{figure4}}
\end{figure}

\clearpage

\begin{figure}[htb]
\begin{center}
\includegraphics[width=16cm]{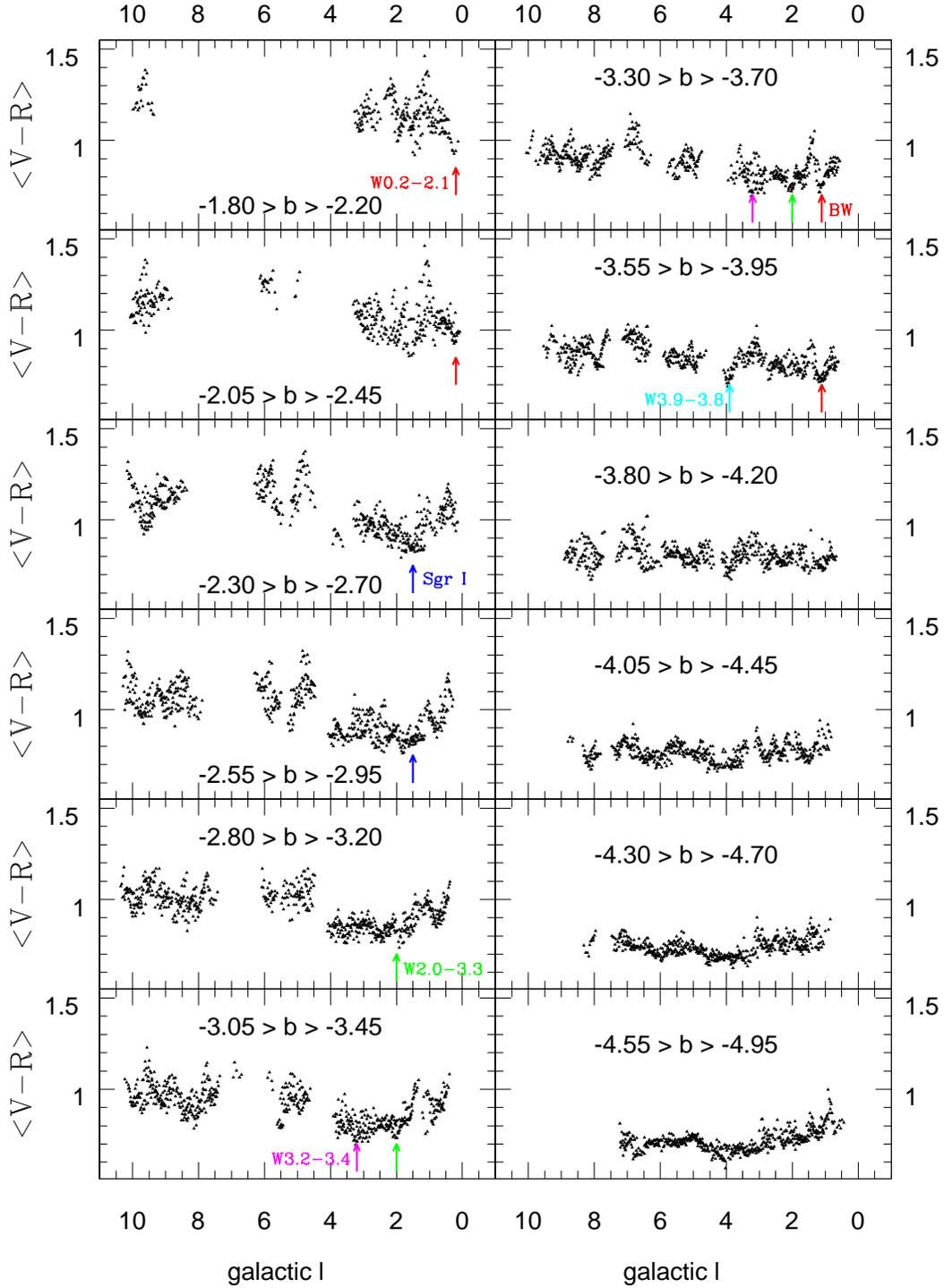}
\end{center}
\vspace*{-1.7cm}
\caption{Color versus galactic longitude $l$ in strips covering 
a range of $0 \degdot 4$ in galactic latitude $b$. Low extinction
windows can be identified as troughs in color that persist for at
least two strips. Only the first occurrence (from up to bottom
and from left to right) is marked with a name -- the remaining ones are
indicated with arrows with identical color. A single possible
low-extinction field
at $(l,b) = (3 \degdot 9,-3 \degdot 8)$ is seen in only one $b$ strip, 
but the signal seems to be very strong.
\label{figure5a}}
\end{figure}

\begin{figure}[htb]
\includegraphics[width=16cm]{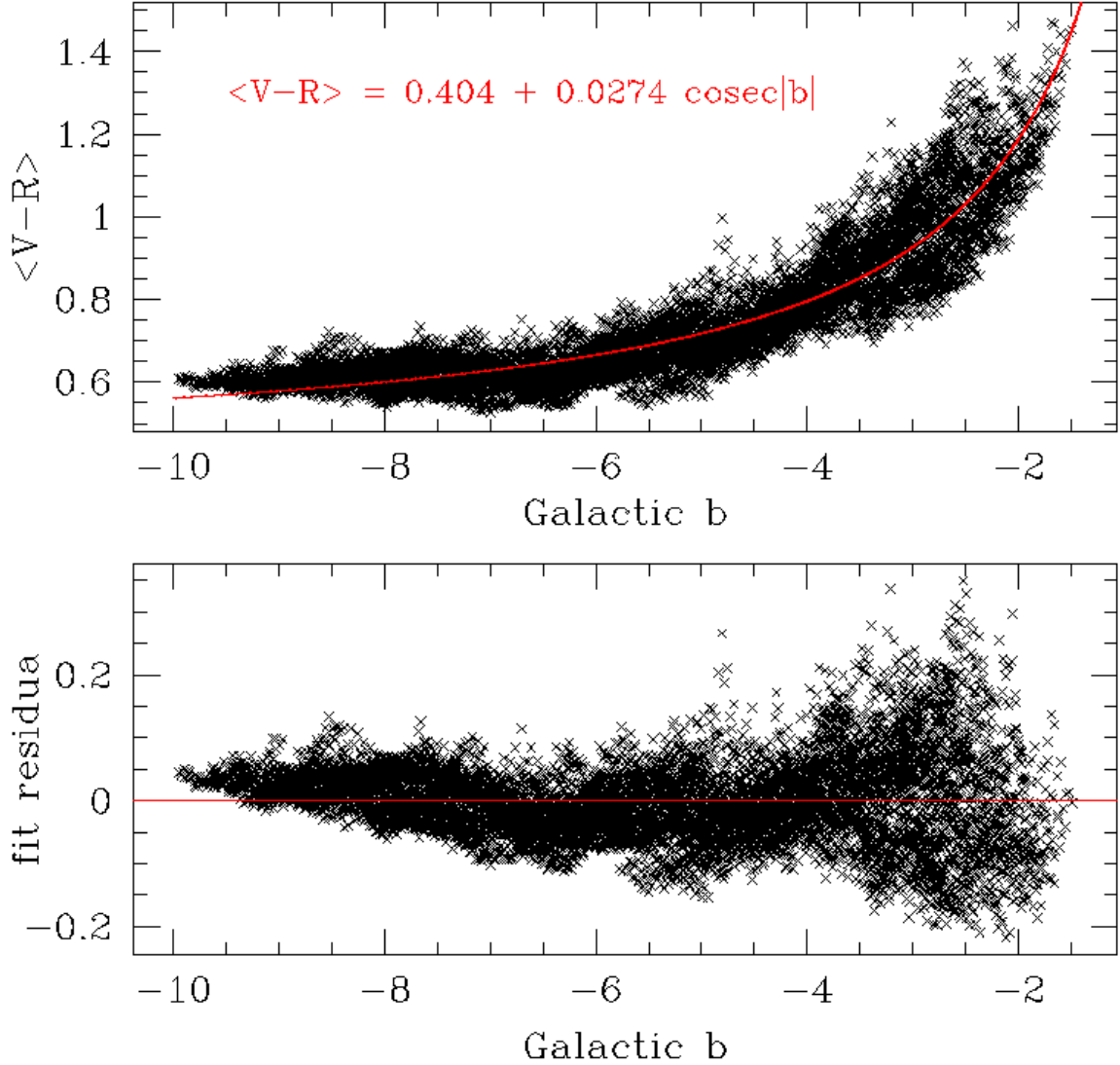}
\caption{The red line displays the best fit of a simple cosec-law model
of dusty disk. The upper panel present the fit, whereas the lower
panel shows the residuals. The fit is based on 9422 resolution
elements with the three disk fields at 
$(l,b) \approx (18^{\circ}, -2^{\circ})$ excluded from the sample.
\label{figure6}}
\end{figure}

\clearpage

\begin{figure}[htb]
\begin{center}
\includegraphics[width=12.5cm]{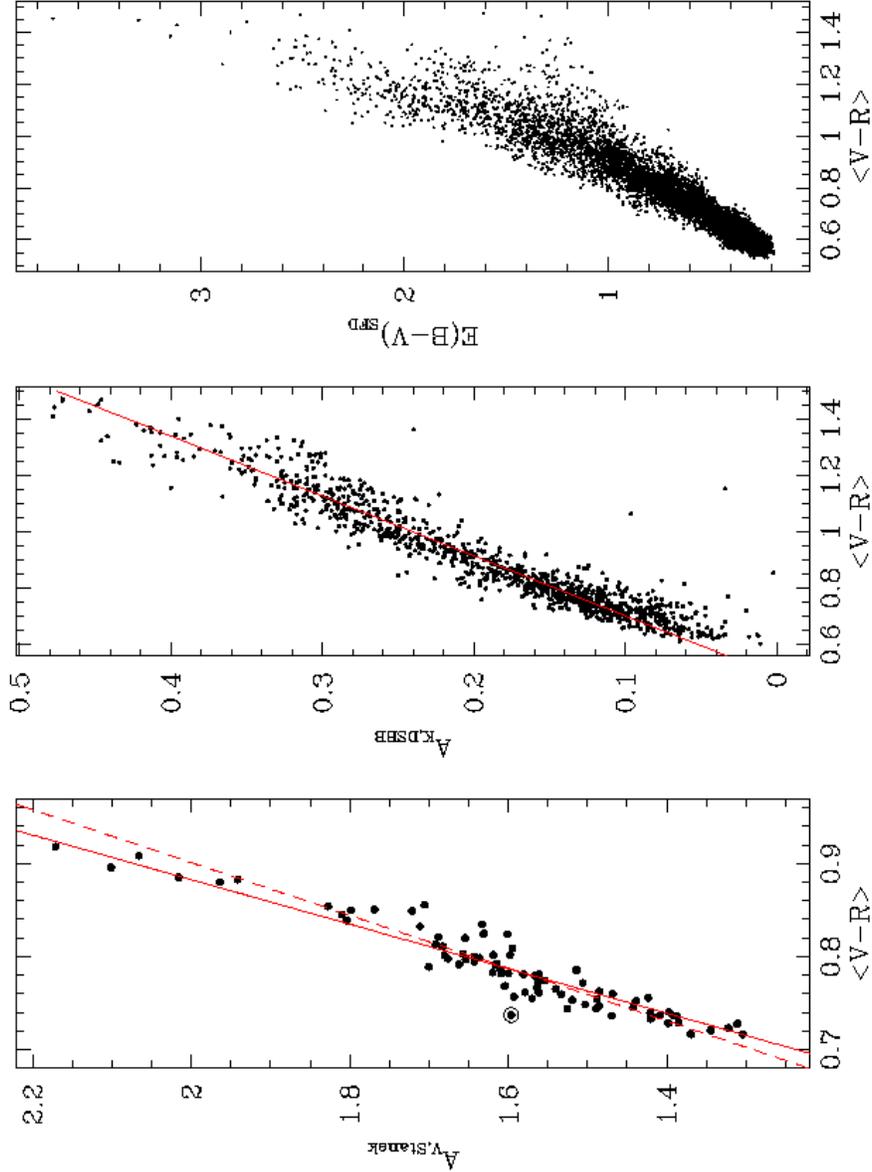}
\end{center}
\caption{The left panel displays
the relation between the mean color determined in this paper
and extinctions obtained by Stanek (1996) for Baade's Window.
As the resolution of our map is lower than that of the OGLE-based
extinction map in
Baade's Window, each $A_V$ value for comparison was obtained by 
averaging between 64 and 81 of Stanek's (1996) 30'' by 30'' resolution 
elements.
The dotted line represent equation\rr{bestfitav} and the solid line
equation\rr{extconv1}. The excluded outlier is marked with a circled point.
The middle panel presents the comparison with Dutra et al.\ (2003)
values, that were based on the analysis of giant branches from 2MASS
data. Here we plot their $A_K$ extinctions versus our $<V-R>$.
We have matches for 1069 resolution elements. The span of $<V-R>$ is
much wider than in the left panel. An over-plotted line comes from the 
scaled equation\rr{jmkfit}.
In the right panel we plot Schlegel et al.\ (1998) reddenings
based on COBE/DIRBE dust emission.
Since the Schlegel et al.\ (1998) reddening map
covers the entire sky, we matched all 9422 tiles from the bulge
region. The relation
between Schlegel et al.\ (1998) reddenings and $<V-R>$ is not linear.
In summary, there is a strong correlation between an average color of the
CMD $<V-R>$ and different reddening/extinction maps.
\label{figure7}}
\end{figure}

\clearpage

\begin{deluxetable}{cc}
\tablecaption{CCD-related correction to $(V-R)$ colors\label{table1}}
\tablewidth{0pt}
\tablehead{
\colhead{CCD designation} & 
\colhead{$(V-R)$ correction [mag]}
}
\startdata 
chip 0 & $-0.0639$\phm{$-$}\\
chip 1 & $0.0000$\\
chip 2 & $-0.0072$\phm{$-$}\\
chip 3 & $-0.0377$\phm{$-$}
\enddata
\end{deluxetable}

\begin{deluxetable}{lccc}
\tabletypesize{\small}
\tablecaption{Low-extinction windows identified in the MACHO bulge data\label{table2}}
\tablewidth{0pt}
\tablehead{
\colhead{Window name} & 
\colhead{Approximate location $(l,b)$} &
\colhead{$<V-R>_{\rm min}$} &
\colhead{Comment}
}
\startdata 
Baade's Window & $(1 \degdot 1,-3 \degdot 9)$ & 0.72 & Baade (1963)\\
Sgr I & $(1 \degdot 5,-2 \degdot 6)$ & 0.81 & Baade (1963)\\
Sgr II & $(4 \degdot 0,-5 \degdot 1)$ & 0.57 & Baade (1963)\\
W0.2-2.1 & $(0 \degdot 2,-2 \degdot 1)$ & 0.93 & Stanek (1998), Dutra et al.\ (2002)\\
W2.0-3.3 & $(2 \degdot 0,-3 \degdot 3)$ & 0.72 & new\\
W3.2-3.4 & $(3 \degdot 2,-3 \degdot 4)$ & 0.70 & new\\
W3.9-3.8 & $(3 \degdot 9,-3 \degdot 8)$ & 0.69 & new\\
W3.2-6.4 & $(3 \degdot 2,-6 \degdot 4)$ & 0.54 & \phm{(?)} new (?)
\enddata
\end{deluxetable}

\begin{deluxetable}{ccccc}
\tabletypesize{\small}
\tablecaption{Fragment of the extinction table\label{table3}}
\tablewidth{0pt}
\tablehead{
\colhead{Galactic $l$} &
\colhead{Galactic $b$} & 
\colhead{$<V-R>$} &
\colhead{$A_V$ [mag]} &
\colhead{$A_V$ [mag]}\\
\colhead{} & \colhead{} & \colhead{} &
\colhead{calibrated to Stanek (1996)} & \colhead{calibrated to Dutra
et al.\ (2003)} 
}
\startdata 
1.142 & $-2.060$ & 1.462 & 4.42 & 4.53 \\
1.143 & $-3.852$ & 0.737 & 1.39 & 1.22 \\
1.143 & $-2.635$ & 0.858 & 1.90 & 1.77 \\
1.144 & $-4.679$ & 0.764 & 1.50 & 1.34 \\
1.145 & $-5.797$ & 0.625 & 0.93 & 0.71 \\
1.148 & $-5.944$ & 0.640 & 0.99 & 0.78 \\
1.149 & $-4.826$ & 0.789 & 1.61 & 1.46 \\
1.149 & $-3.999$ & 0.744 & 1.42 & 1.25 \\
1.151 & $-6.090$ & 0.592 & 0.79 & 0.56 \\
1.151 & $-2.208$ & 1.108 & 2.94 & 2.91 \\
1.153 & $-6.237$ & 0.576 & 0.72 & 0.49 \\
1.154 & $-4.973$ & 0.765 & 1.51 & 1.35 \\
1.155 & $-4.147$ & 0.785 & 1.60 & 1.44 \\
1.155 & $-3.466$ & 0.753 & 1.46 & 1.29 \\
1.156 & $-6.383$ & 0.593 & 0.79 & 0.57 \\
1.157 & $-1.821$ & 1.250 & 3.53 & 3.56 \\
\enddata
\end{deluxetable}


\clearpage

\begin{references}
\reference{alc1} Alcock, C., et al.\ 1995, \apj, 445, 133
\reference{alc2} Alcock, C., et al.\ 1997, \apj, 474, 217
\reference{alc3} Alcock, C., et al.\ 1998, \apj, 494, 396
\reference{alc4} Alcock, C., et al.\ 1999, \pasp, 111, 1539
\reference{alc5} Alcock, C., et al.\ 2000, \apj, 541, 734
\reference{arc} Arce, H.C., \& Goodman, A.A. 1999, \apj, L135
\reference{baa} Baade, W. 1963, Evolution of Stars and Galaxies,
Harvard Univ. Press, Cambridge MA, p. 277
\reference{car} Cardelli, J.A., Clayton, G.C., \& Mathis, J.S. 1989,
\apj, 345, 245
\reference{dut1} Dutra, C.M., Santiago, B.X., Bica, E. 2002, \aap, 381, 219
\reference{dut2} Dutra, C.M., Santiago, B.X., Bica, E.L.D., \& Barbuy,
B. 2003, \mnras, 338, 253 (DSBB)
\reference{fro} Frogel, J.A., Tiede, G.P., \& Kuchinski, L.E. 1999,
\aj, 117, 2296 
\reference{gir} Girardi, L., Bertelli, G., Bressan, A., Chiosi, C.,
Groenewegen, M. A. T., Marigo, P., Salasnich, B., Weiss, A. 2002, 
\aap, 391, 195
\reference{gou} Gould, A., Popowski, P., \& Terndrup, D.M. 1998, \apj,
492, 778
\reference{kir1} Kiraga, M., \& Paczy\'{n}ski, B. 1994, \apj, 430, L101 
\reference{kir2} Kiraga, M., Paczy\'{n}ski, B., \& Stanek, K.Z. 1997,
\apj, 485, 611
\reference{knu} Knude, J. 1996, \aap, 306, 108
\reference{odo} O'Donnell, J.E. 1994, \apj, 422, 158
\reference{pop} Popowski, P. 2000, \apj, 528, L9
\reference{schl} Schlegel, D.J., Finkbeiner, D.P., \& Davis, M. 1998,
\apj, 500, 525 (SFD)
\reference{schu} Schultheis, M., et al.\ 1999, \aap, 349, 69
\reference{sta} Stanek, K.Z. 1996, \apj, 460, L37
\reference{sta} Stanek, K.Z. 1998, preprint (astro-ph/9802307)
\reference{sum} Sumi, T., et al.\ 2003, \apj, 591, 204
\reference{uda} Udalski, A. 2003, \apj, 590, 284
\reference{woz} Wo\'{z}niak, P.R., \& Stanek, K.Z. 1996, \apj, 464, 233
\end{references}
\end{document}